\documentstyle[version2,preprint,aps]{revtex} %preprint
\begin{document}

\preprint
{
\begin{flushright}
LAEFF--93/017\\
December 1993
\end{flushright}
}

\vspace{-.25in}

\begin{title}
The Renormalization Group, \\
Entropy, Thermodynamic Phase Transitions and Order\\
in\\
Quantum Field Theory.\\
\end{title}

\author{J.\ P\'erez--Mercader\cite{AAAuth}}
\begin{instit}

Laboratorio de Astrof\'{\i}sica Espacial y F\'{\i}sica
Fundamental\\
Apartado 50727\\
28080 Madrid
\end{instit}

\vspace{-.125in}
\begin{center} \bf
{\sl Submitted to Physics Letters B}

\end{center}
\vspace{-.125in}
%\begin{center}
%December 3, 1993
%\end{center}

\vspace{-.25in}

\begin{abstract}

We define an entropy for a quantum field theory by combining quantum

fluctuations, scaling and the maximum entropy
concept. This entropy has different behavior in asymptotically free
and non--asymptotically free theories. We find that the transition
between the two regimes (from the asymptotically free to the
non--asymptotically free) takes place via a continuous phase
transition. For asymptotically free theories there
exist regimes where the ``temperatures" are negative. In
asymptotically free theories there exist maser--like states mostly
in the infrared; furthermore, as one goes into the ultraviolet and
more matter states contribute to quantum processes, the quantum
field system can shed entropy and cause the formation of
thermodynamically stable {\it entropy--ordered} states. It is

shown how the known heavier quarks can be thus described.

\end{abstract}

Quantum fluctuations are an unavoidable feature of any quantum
theory,

and quantum field theory in particular. They modify physical
quantities,

which become scale dependent objects, in a way which reflects the

quantum nature of the underlying virtual cloud. At a given scale,
where

quantum fluctuations are active, there are contributions from virtual

processes with an arbitrary (but smaller than this scale) impact

parameter. In particular, they change the interaction energy which
now

has an extra, ``built--in", indeterminacy, manifesting as a deviation

from its $classical$ form. These corrections are taken into account
in

quantum field theory via the renormalization process, and their scale

dependence is quantitatively described by the renormalization group

equations (RGEs) satisfied by the corresponding $n$--point

functions \cite{gellmannlow}.

Another general feature of fluctuating systems is that one can
describe

many interesting properties of their collective behavior by
associating

with the system some probability distribution. From this probability

distribution one can calculate, e.g., the entropy associated with it
and

obtain information on structural properties of the system.

We will introduce such a density for a quantum field system in its

static limit. For this, we take advantage of the connection between

potential theory and probability theory \cite{doob} where, for a

potential that satisfies a Poisson equation, one can interpret the
associated density

as a probability density.

The starting point for our purposes is the quantum corrected, static

limit of the interaction energy, which satisfies a Poisson equation,

and that we can use to write down the probability density which will

allow us to study some order-disorder properties of the quantum
system.

At short distances the $leading$ term in the static interaction
energy between elementary ``charges" interacting in the quantum
vacuum,

and for massless mediating
quanta, can be written at one loop, as \cite{self}

\begin{equation}
V(r) \cong \frac{C}{4 \pi}g_0^2 r^{-1+\sigma}
\label{potential}
\end{equation}

\noindent

where,  $\sigma$ is related to the beta function for the
coupling $g_0$ by $\sigma=+ 2 \beta_0 g_0^2$, and $\mu d g_0^2/d \mu
=-\beta_0 g_0^4$, with $\mu$ the momentum scale.

Because of the Poisson equation satisfied by $V(r)$, away from $r=0$,

there exists \cite{doob} a density $\rho (r) \cong
r^{-3+\sigma}$ which we interpret as a probability density (in the

classical sense of probability) for the distribution of the virtual

cloud that surrounds the elementary charges in the quantum-mechanical

vacuum. Normalization of this charge density is possible if one

introduces an IR-cut-off $R_0$ in the case of asymptotically free
(AF)

theories and an UV-cut off $r_0$ for non-asymptotically free (NAF)

theories. One gets

\begin{equation}
\rho (r)=Ar^{-3+\sigma }
\label{density}
\end{equation}

\noindent
where $A={\sigma  \over {4\pi}}R_0^{-\sigma }$ when $\sigma >0$ and

$A={{-\sigma }\over {4\pi }}r_0^{+\sigma}$ if the theory is
NAF. \footnote{It is interesting to mention that this probability
density is of the type associated with Lotka's or Zipf's laws, well
known as typical examples of Pareto--Levy distributions
\cite{montroll}.}

The fine grained (Gibbs') entropy for $\rho$ can be computed directly

from the definition $S=-\sum_i p_i \ln p_i$,
where $p_i$ is a probability density ($\rho \equiv p_i$).
Alternatively,

one may subject the system
\cite{montroll} to the constraint $\left\langle f
\right\rangle=\sum\limits_{i=1}^n {p_i\,f(x_i)}$, and perform a

variational calculation where one is (formally) led to

\begin{equation}
p_i={{e^{-\beta f(x_i)}} \over {Z(\beta )}}, \,\,\,\,\,
Z(\beta )=\sum\limits_{i} {\exp \left[ {-\beta f(x_i)}
\right],} \label{partition}
\end{equation}

\noindent
and the entropy can be recast as

$$
S=\log Z(\beta )+\beta \sum\limits_i f(x_i)
\frac{\exp \left[ -\beta f(x_i)\right]}{Z(\beta )}
$$
\begin{equation}
\equiv \log Z( \beta )+\beta U
\label{entropy}
\end{equation}

\noindent
{}From this equation, we see that $U$ plays the r\^ole of an internal

energy, whereas $1/\beta$ can be
identified with the ``temperature" $T$ of the virtual cloud.
\footnote{From now on, we will drop the quotes around
``temperature". The reader is kindly asked to mentally replace them
each time she/he sees the word temperature in this paper. We will
adopt ``natural" units and take the equivalent of Boltzmann's
constant equal to 1. This $\beta$ should not be confused with
$\beta_0$,

the RG--quantity introduced above.}

Using these formulae on the density of Eq. (\ref{density}), we

perform at once the following identifications:

\begin{equation}
\beta = 3-\sigma; \,\,\,\, f(r) =\log r/R_0; \,\,\,\,
Z(\beta)=\frac{4 \pi}{\sigma {\Theta}} R_0^3
\label{temperature}
\end{equation}

\noindent
for AF--theories, and for NAF--theories

\begin{equation}
\beta = 3+|\sigma|; \,\,\,\, f(r) =\log r/r_0; \,\,\,\,
Z(\beta)=\frac{4 \pi}{|\sigma| {\Theta}} r_0^3
\label{temperatureNAF}
\end{equation}

\noindent
Here ${\Theta}$ plays a r\^ole similar to the
``entropy constant" which is fixed in quantum statistical mechanics
by using Nernst theorem. In principle, ${\Theta}$ is an $arbitrary$
quantity with dimension of length to the cube.

We see at once

that $\sigma$ is related to the inverse temperature introduced in
the variational calculation. The temperature ranges are as follows:
$0 \leq T \leq 1/3$ for NAF--theories; in AF--theories, $T$ must be
$\geq 1/3$ or $0 \geq T$, so that, in principle, negative
temperature regimes are allowed in these theories. In Figure 1, we
show the entropy as a function of $T$ for both AF-- and
NAF--theories.

The energy $U$ is given (for both NAF-- and AF--theories) by

\begin{equation}
U=\frac{T}{1-3T}
%+ \log \left( \frac{R_0}{(C_N)^T} \right)
\label{energyAF}
\end{equation}

\noindent
With these identifications, the entropy obtained from
Eq.(\ref{entropy}) coincides with the fine--grained entropy
calculated from the probability density of Eq.(\ref{density})
\cite{self}.

The specific heat at constant volume $C_V$ (again for both types of
theories) is given by

\begin{equation}
C_V=\left . \frac{\partial U}{\partial T}\right|_V=\frac{1}{(1-3T)^2}
%- \log C_N
\label{cv}
\end{equation}

The specific heat blows--up (cf. Figure 2) when $T= 1/3$. A phase
transition takes place in the non--abelian $and$ static quantum
system when $\sigma =0$, that is when the system goes from the
non--abelian to the abelian phase.\footnote{This seems to also
indicate the following: at the classical limit ($\hbar \rightarrow
0$) $\sigma$, which is proportional to $\hbar$, goes to zero; thus a
phase transition of this kind can take place when a quantum field
theory (both AF and NAF) goes into its classical regime.}  As is
seen from (\ref{cv}), $all$ derivatives of $U$ diverge at $T=1/3$,
and
the transition is a continuous phase transition. For an AF--theory
the
transition occurs as the size of the probed system decreases and, due
to
the decoupling theorem \cite{decoupling}, more matter degrees of
freedom
start contributing to the beta--function, which has the possibility
of
changing sign and become negative.\footnote{This change of phases in
a
quantum field theory is $unrelated$ to the magnetostatic
classification
of the vacuum based on renormalization group behavior given by Pagels
and
Tomboulis in Reference \cite{pagelsandtomboulis}. The phase
transition described here requires a change in the sign of $\sigma$
$or$ that $\sigma$ be zero; the former is possible only for
AF--theories, whereas the latter is also true for NAF--theories in
their transition to the classical regime, as already pointed out.}

We now analyze some features specific to AF--systems. In order to do
this
we introduce the variable $x \equiv -1/T$, which has the property
that
colder temperatures are mapped to $- \infty$, hotter to $+ \infty$
and is useful in the description of negative temperature
systems \cite{ramsey}. For AF--regimes $x$ ranges between $x=-3^+$
and $x= + \infty$ (cf. Figure 3). In Figure 4, we plot $S$, $U$ and
$C_V$ as functions of $x$; we see that $C_V$ has a minimum and
vanishes at $x=0$ $(\sigma =+3)$ while the entropy has a maximum
there.

The entropy of the system at first increases as we go from $x=-3^+$
to $x=0^-$
(which corresponds to a positive increase in $\sigma$) to then
decrease from
$x=0^+$ to $+\infty$. Since $\sigma$ becomes more and more positive
as we go up
in the size of the system we probe (less matter degrees of freedom
contributing
to $\beta_0$), we see that the entropy at first increases from $-
\infty$ at
$x=-3$ to its maximum at $x=0$, to then $decrease$ with increasing
system size:
thus when $\sigma=3$ there is some form of ``condensation" (or
population
inversion) that sets in.\footnote{As may be seen from Equations
(\ref{potential}) and (\ref{density}), this happens when $\sigma
\equiv$
dimensionality of the physical $space$.}

The specific heat decreases very abruptly after the phase transition
at
$T=1/3$, (accompanied by an equally abrupt increase in the entropy)
and becomes
zero at $\sigma =3$. For $\sigma >3$, the system is in a {\it totally
new
state}, where the specific heat increases with $\sigma$ and tends
asymptotically to +1, as the size of the probed system is increased;
that is,
as we try to ``raise the temperature" the system absorbs ``energy"
(up to a
maximum of $U(\sigma=\infty)-U(\sigma=3) = +1/3$) coming to a state
where a
large increase in negative temperature (for $\sigma >>3$) increases
the
internal energy very little and yet the system absorbs it (the energy
goes
asymptotically to zero as $C_V \rightarrow +1$). Since this  is
accompanied by

a decrease in entropy, one is tempted to interpret this situation as
an
indication that quantum field theory favors the existence of
$organized$ and
complex states when $\sigma$ becomes $>3$. This observation is
prompted by the
notion that a system sheds entropy to increase its level of
organization \cite{nature} or complexity \cite{bennett}.
Unfortunately,
for quantum chromodynamics (QCD), $this$ range is well inside the
non--perturbative regime where one has to question the validity of
the
perturbative calculations that led to Eq.(\ref{density}), but it is
present in the perturbative regime of other gauge theories.

Similar considerations apply when $\sigma$ is less than 3 and as it
approaches zero, during the infrared--to--ultraviolet transition. In
principle, within this range, the entropy decreases with $\sigma$,
and the quantum system has the potential for sustaining the formation
of organized states at a lower $\sigma$ by shedding entropy as the
size
of the probed region is decreased when going from the infrared to the
ultraviolet. At $each$ value of $\sigma$, there is a $unique$ value
of
$\Theta$ that makes the entropy equal to zero, $and$ leads to
thermodynamical equilibrium ($\partial A / \partial T \left.
\right|_V =
-S(\sigma, \Theta, R_0)=0$, where $A$ is the free energy). Writing

$\Theta =  \bar{r}^3 \exp a$, where $a$ is an unspecified constant
and
$\bar{r}$ the  maximum system
size that can be (adiabatically) supported for this value of
$\sigma$, the above condition can only be satisfied for a pair of
values $(\sigma, R_0/ \bar{r})$ once $a$ has been $fixed$ at, say,
$\bar{r}=R_0$. It is very illustrative to look at this situation from
the  point of view of order--disorder. For $\sigma < 3$, when we
probe
the  system at shorter and shorter distances, the system ``cools",
and
the  entropy decreases (cf. Figs. 1 and 4); in other words, going
into
the  UV is accompanied by a drop in the temperature, which is

a ``disorder--decreasing" process, reflected in QCD by a decrease

in the entropy. But in quantum field theory, as we decrease the size

of the system being probed, more momentum transfer is available,
$new$

particle thresholds are crossed and $more$ degrees of freedom become

active in the quantum system; the presence of additional degrees of
freedom at shorter distances raises the number of states accessible
to the
system,  a ``disorder--increasing" process which makes the entropy
grow.
The  $simplest$ possibility for the $net$ entropy is to assume that
these

two competing effects compensate each other, and that the process of

probing the shorter distances is isentropic.

The isentropic nature of this process implies that the entropy

\begin{equation}
S^{(\sigma > 0)}=1-\frac{3}{\sigma} -\ln \frac{\sigma}{4\pi}
+ 3 \ln \frac{R_0}{\bar{r}} -a
\label{3}
\end{equation}

\noindent
stays constant, and the excess entropy goes into
$organizing$ and $supporting$ new states. In QCD, where $\beta_0$
decreases each time the threshold for a new flavor is crossed, this
insinuates the existence of a relationship between the different
quark states, in  particular, their masses must be related according
to
Eq.(\ref{3}) with $\Delta S=0$ between flavors. This  relationship is
shown in Figure 5. We compare the known experimental heavier quark
masses
(i.e., all except $up$ and $down$) and strong interaction coupling
constant, with the $result$ of $assuming$ $\Delta S=0$ between quark
states. The experimental data can be described by adjusting $one$
single
free parameter, the zero of the entropy: the resulting correlation
between
theory and experiment is better than .99. In this figure we have
fitted
$\Delta S=0$, by adjusting $a$ of Eq.(\ref{3}), to the known
experimental
data \cite{rpp}: {\it the product is that the theoretical curve is
never
more than 8\% away from the central experimental values}.

The nature of the organized states for $\sigma >3$ and $\sigma <3$ is

quite different: while the states with $\sigma >3$ are
thermodynamically unstable, the others are thermodynamically stable

states. This follows from  Equation (\ref{cv}) and the second
derivative of the free energy with respect to the temperature,

\begin{equation} \left. \frac{\partial^2 A}{\partial T^2}\right|_V =
-\left. \frac{\partial S}{\partial T}\right|_V=-\frac{C_V}{T}
\label{stability} \end{equation}

\noindent In thermodynamically stable states this quantity must be
negative \cite{reichl}. We see that the states in the region of
negative temperatures are un--stable because they correspond to a
free energy which is a convex function of the temperature; the
converse is true for the ``entropic states" in the region of
positive temperatures.

In summary, quantum field theories, through quantum fluctuations,
as described by the renormalization group, and in the leading
approximation, have charge densities which can be interpreted as

probability densities of the class associated with Pareto--Levy
\cite{montroll} distributions. This leads to a generalized notion of
entropy and ``temperature" in quantum field theory. For AF--theories
there exist regimes of negative temperatures. At the transition from
the AF--regime to the NAF--regime, there is a continuous phase
transition. In AF--theories and when $\sigma \geq 3$ there is the
possibility that there exist organized, highly complex, albeit
thermodynamically unstable, maser--like states. However, in the
regime where $\sigma \leq 3$, AF--theories support the existence of
thermodynamically--stable--states whose order is provided by
entropic effects; these states become patent as the quantum system
goes from the infrared into the ultraviolet and the system entropy
falls. Finally, the ideas discussed here and the results found are
very general, and they can find application for example in grand

unification physics, the quantum theory of gravity, multiparticle
physics, astrophysics, the very early universe and those realms of
physics where quantum field theory and/or the renormalization group
are applicable, including fractal dynamics, chaos and complexity
theory.

\acknowledgements
I have benefitted from discussions with M. Bastero,  R.
Blankenbecler,
T. Goldman, S. Habib, I. M. Khalatnikov and G. West.

\figure{The entropy of asymptotically free and non--asymptotically
free theories. The zeroes of the entropies have been (arbitrarily)
chosen so as to display the curves in a position where their features
can be easily appreciated.}

\figure{$C_V$ and $\sigma$ plotted as a function of the
``temperature".
 The quantity $\sigma$ increases with the size of the region of space
being probed.}

\figure{The ``temperature" $T$, and the variable $x$ plotted against
the renormalization group quantity $\sigma$. The quantity $\sigma$
increases with the size of the region of space being probed.}

\figure{Entropy, $C_V$ and internal energy as functions of the
variable
$x=-1/T$. The variable $x$ increases with the size of the region
of space being probed. The zero of the entropy has been (arbitrarily)
chosen so as to display the curves in a position where their
different features can be easily appreciated.}

\figure{The condition $\Delta S =0$ for quarks in QCD. The
experimental data have been fitted setting $S^{(\sigma>0)}
(\sigma)$ of Equation (\ref{3}) equal to zero, with
(cf. Eq. (\ref{3})) $a=-0.721151$. The experimental values are
given in Table 1. The Pearson's correlation coefficient between
experimental data and theory is greater than 0.99.
In the inset we plot the residues for this fit. The reduced
$\chi^2$ is 0.1120. All known quark masses are fitted with an error
of
less than 8\% .}

\newpage

\mediumtext
\begin{table}
\setdec 00.000000
\caption{Experimental values used in fit for the quark masses.$^{\rm
a}$}
\begin{tabular}{c|c|c|c}
Flavor/State & $<3\ \ln (m_f/2 m_{strange})>$
& $<\sigma_{exp}>$ & Residue \\
\tableline
$b\bar{b}$ &\dec 8.440232 &\dec 0.206233 &\dec -0.27553 \\
$b$ &\dec 6.360791 &\dec 0.240759 &\dec -0.42370 \\
$c\bar{c}$ &\dec 4.828314 &\dec 0.302973 &\dec 0.37270 \\
$c$ &\dec 2.748872 &\dec 0.384270 &\dec 0.15042 \\
$s\bar{s}$ &\dec 0.0 &\dec 0.670211 &\dec 0.17611 \\
\end{tabular}
\tablenotes{$^{\rm a}$ $\alpha_{strong} (M_Z)= 0.116 \pm 0.005$.
The scale for $2 m_{strange}$ was taken at 0.6 $GeV$; the
data on quark masses is from Ref. \cite{rpp}. In addition to the
quark
masses, we have also included the masses of the corresponding
$q\bar{q}$--states.}

\end{table}

\end{document}